\documentclass[aps,pre,preprint,floatfix]{revtex4}

\usepackage{latexsym}
\usepackage[usenames]{color}
\usepackage{amsthm}
\usepackage{hyperref}

\usepackage{amsmath}    
\usepackage{amsfonts}  
\usepackage{amssymb}
\usepackage{graphicx}
\usepackage{slashed}
\usepackage{fouridx}

\theoremstyle{plain} 
\theoremstyle{plain} 
\theoremstyle{plain}  
\theoremstyle{plain} 
\theoremstyle{plain}  
\theoremstyle{remark} 
\theoremstyle{plain} 
\theoremstyle{remark}

\newcommand\CROSS[1]{%
  \hbox{%
    \vbox{
      \hrule
      \kern2.5pt
      \hbox{$#1$\,\strut}
    }%
  \vrule
  }\mskip\thickmuskip
}

\begin{document}
\begin{center}
\Large{\textbf{From deterministic cellular automata \\ to coupled map lattices}}\\ 
~\\

\large{Vladimir Garc\'{\i}a-Morales}\\

\normalsize{}
~\\

Departament de Termodin\`amica, Universitat de Val\`encia, \\ E-46100 Burjassot, Spain
\\ garmovla@uv.es
\end{center}
\small{A general mathematical method is presented for the systematic construction of coupled map lattices (CMLs) out of deterministic cellular automata (CAs). The entire CA rule space is addressed by means of a universal map for CAs that we have recently derived and that is not dependent on any freely adjustable parameters. The CMLs thus constructed  are termed real-valued deterministic cellular automata (RDCA) and encompass all deterministic CAs in rule space in the asymptotic limit $\kappa \to 0$ of a continuous parameter $\kappa$. Thus, RDCAs generalize CAs in such a way that they constitute CMLs when $\kappa$ is finite and nonvanishing. In the limit $\kappa \to \infty$ all RDCAs are shown to exhibit a global homogeneous fixed-point that attracts all initial conditions. A new bifurcation is discovered for RDCAs and its location is exactly determined from the linear stability analysis of the global quiescent state. In this bifurcation, fuzziness gradually begins to intrude in a purely deterministic CA-like dynamics. The mathematical method presented allows to get insight in some highly nontrivial behavior found after the bifurcation.}
\noindent  
~\\

\pagebreak


\section{Introduction}

In the quest to understand the spatiotemporal dynamics of nonlinear systems with many degrees of freedom, of interest to physics and biology, a hierarchy of different classes of mathematical models has been advanced \cite{Bunimovich1}. On the bottom of the hierarchy, there is the fully discrete dynamics, as described by cellular automata (CAs) \cite{Wolfram, Chua1, Ilachinski, Adamatzky, McIntosh, Wuensche, Ceccherini, VGM1, VGM2}, in which both time and space are discrete and in which the local phase space is discrete and finite (and hence, usually restricted to a finite set of integers). The next upper level in the hierarchy corresponds to coupled map lattices (CMLs) \cite{Kaneko1, Kaneko2, Kaneko3, Kaneko4, Kapral, KanekoCrutch, KanekoBOOK, Bunimovich2, Bunimovich3} in which time and space are discrete as in CAs, but the local phase space is continuous (the local dynamical variable being now real valued). Finally, the uppermost level is constituted by (nonlinear) partial differential equations \cite{Hohenberg, Toffoli, Omohundro} in which all, time, space and local phase space are continuous. All these models constitute different approaches to spatiotemporal pattern formation outside of equilibrium, as found in experimental physical and biological systems \cite{DeutschBOOK}.

Despite the qualitative differences among the classes of models, some spatiotemporal patterns they describe possess strikingly similar features. For example, spatiotemporal intermittency, found in experimental systems \cite{Ciliberto, Daviaud, Bottin, Degen, Colovas, Goharzadeh,  Michalland}, has been modeled with nonlinear partial differential equations, as the complex Ginzburg-Landau equation \cite{Chate4} (see \cite{KuramotoBOOK, Aranson, contemphys} for an introduction), as well as with CMLs \cite{Chate1, Chate2, Chate3, Gupte, Oono1, Oono2} and CAs \cite{Wolfram, Jabeen1, Jabeen2, VGM3}. There are a few articles in which the study of the relationship between CMLs and CAs has been explored \cite{Chate1, Chate2, Chate3}. In most of these seminal works, resort to approximations has been made to get insight in the relationship between some specific CAs and CMLs \cite{Chate1, Chate2, Chate3} or CAs have been obtained as a result of coarse-graining of CMLs \cite{Jabeen1, Jabeen2}. Although some deep results have been obtained (for example an elegant means to construct Class 4 CA from CMLs \cite{Chate2, Chate3}) these are specific to some particular CAs and the CA rule space was not systematically addressed as a whole. The problem of how to preserve certain  continuum properties as isotropy has been considered by Nishiyama and Tokihiro \cite{Tokihiro6} who construct an isotropic CA model for reaction diffusion systems by exploiting the fact that the continuum limit of a random walk yields a diffusion equation. Very recently \cite{EPL}, we have also proposed a general method to derive CA approximations (shadowings) from CMLs of nonlinearly coupled oscillators.

The study of the interplay between CAs and CMLs is of great interest to the field of discrete integrable systems, where the problem of construction of discretizations preserving some properties (related to integrability) is of fundamental importance. In a pioneering work, Tokihiro et al. \cite{Tokihiro1} established a direct connection between integrable CA and certain nonlinear wave equations, as the Korteweg-de Vries equation, through a general limiting procedure. This method, also called ultradiscretization \cite{Tokihiro2, Tokihiro3, Tokihiro4, Tokihiro5}, allowed the construction of  integrable CAs out of difference-difference equations. In subsequent work, the inverse problem \cite{Tokihiro5} was also addressed. Another method to construct integrable CAs out of complex-valued discrete systems was formulated in \cite{Bialecki1, Bialecki2} and consists in keeping the map of the discrete system unchanged, but changing the underlying field of complex numbers to a finite (Galois) field $\mathbb{F}_{q}$ (where $q$ is a certain power of a prime number $p$). This algebro-geometric construction is obtained starting from nonsingular algebraic curves \cite{Bialecki1, Bialecki2}.

Recently, a parameter-free universal map for CA has been derived \cite{VGM1} which allows the whole class of deterministic CA to be mathematically handled. We believe that, because of CAs being `simpler' than CMLs, it might also be a good strategy to approach the study of complex nonlinear behavior starting from this lowest level in the hierarchy (namely CAs) and then proceeding to exactly derive CMLs from them in a bottom-top approach. Looking for  natural embeddings of all finite sets of integer numbers in the field of real numbers suggests a pathway to carry out this program. 

In this article, we follow this strategy tackling the problem of establishing a wide family of CMLs that \emph{generalize} CAs by encompassing all them as the proper limit ($\kappa \to 0$) of a certain one-parameter family of functions $\mathcal{B}_{\kappa}$. All main results are exact and allow a new bifurcation in the passage from deterministic CA-like behavior to CML-like behavior to be uncovered. Starting from the universal map for deterministic CAs \cite{VGM1}, we embed this universal class of dynamical systems in a specific class of CMLs that explicitly depends on the continuous parameter $\kappa$ and that we call \emph{real-valued deterministic CA} (RDCA) such that in the asymptotic limit $\kappa \to 0$, the deterministic CAs used in the construction are regained. Thus, in brief, RDCAs are CMLs that arise from an embedding of CAs on a continuum local phase space. Although our starting point in the construction is the class of all 1D deterministic CA which are first-order in time, the method can be extended in a straightforward manner to \emph{any} deterministic CA in any dimension and any order in time, considering the appropriate maps for these situations \cite{VGM1}. 

The outline of this article is as follows. In Section \ref{CA2CML} we present the construction of the RDCA from our previously derived universal map for deterministic CA. In this new concept, CMLs and deterministic CAs merge together in an unified manner. We then present some main features of RDCAs, which lead to uncover a new bifurcation between deterministic CA computation and a more fuzzy CML-like behavior. An explicit expression for the location of the bifurcation in parameter space is determined through the linear stability analysis of the global homogeneous mode for all RDCA which fix the quiescent state in the asymptotic limit $\kappa \to 0$ (called the CA limit). In Section \ref{numerical} with help of some computer explorations, we confirm our analytical results and explore some nontrivial dynamical phenomena, as well as the impact of the bifurcation. Finally, we uncover a new kind of spatiotemporal intermittency for some RDCA rules. We then summarize some conclusions of this study.

\section{Real-valued deterministic cellular automata (RDCA)} \label{CA2CML}

\begin{figure}
\includegraphics[width=0.85 \textwidth]{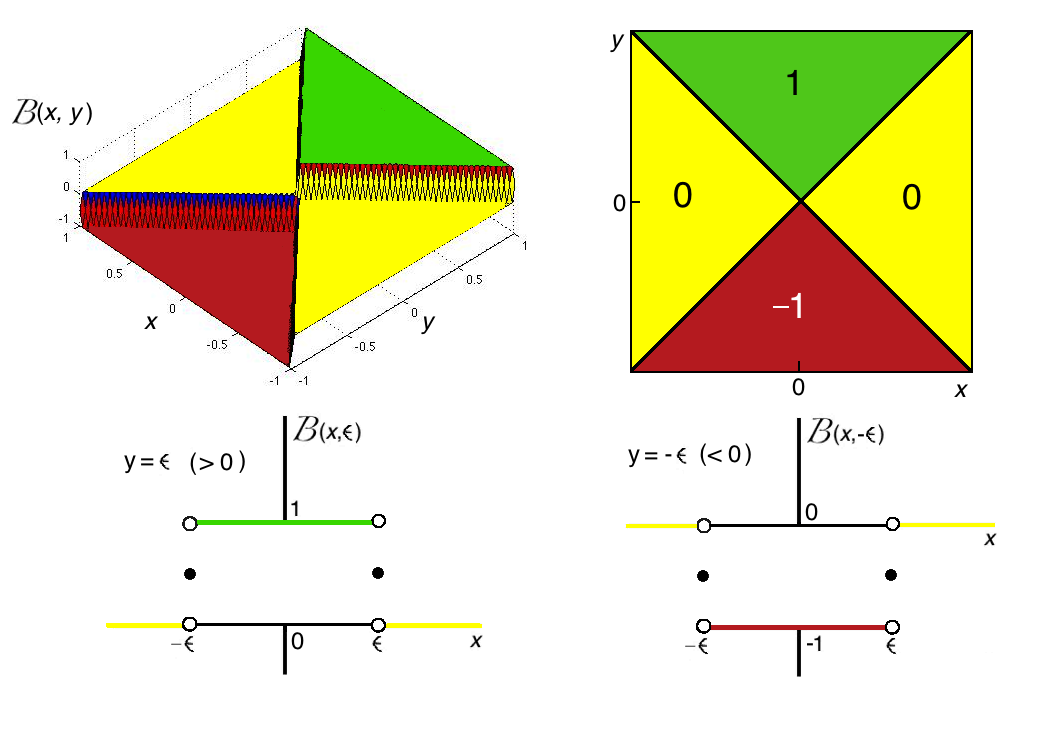}
\caption{\scriptsize{Two views of the $\mathcal{B}$-function (top panels) and for $y=\pm \epsilon$ constant in the second argument (bottom panels).}} \label{box}
\end{figure}

In this work, we denote by $S$ the set of integers $\in [0,p-1]$, where $p \ge 2 \in \mathbb{N}$. We consider a CA dynamics on a 1D ring of $N_{s}$ sites. A CA is a fully discrete system whose \emph{global} dynamics is given by a map $S^{N_{s}} \to S^{N_{s}}$, i.e. $(x_{t}^{0},...,x_{t}^{N_{s}-1}) \to (x_{t+1}^{0},...,x_{t+1}^{N_{s}-1})$ where $x_{t}^{j} \in S$ is the dynamical state of the site at position $j \in \mathbb{Z}$  $\mod N_{s}$ at time $t \in \mathbb{Z}$. The \emph{local} dynamics of the CA is explicitly given by the universal CA map \cite{VGM1, VGM2} as
\begin{equation}
x_{t+1}^{j}=\sum_{n=0}^{p^{r+l+1}-1}a_{n}\mathcal{B}\left(n-\sum_{k=-r}^{l}p^{k+r}x_{t}^{j+k},\frac{1}{2}\right) \qquad j=0,1,\ldots N_{s}-1 \label{CA}
\end{equation} 
where $l, r \in \mathbb{N}$ specify the spatial range of the interactions, the neighborhood of a site $j$ having size $\rho=l+r+1$. The coefficients $a_{n} \in S$ specify the output $x_{t+1}^{j}$ of the CA rule (for example, Wolfram's rule 110, has $p=2$ rule vector $(a_{0},a_{1},...a_{7})=(0,1,1,1,0,1,1,0)$ and $l=r=1$). Note that in Eq. (\ref{CA}) we have $x_{t+1}^{j}=a_{m}$ if and only if $\sum_{k=-r}^{l}p^{k+r}x_{t}^{j+k}=m$ at time $t$. In Eq. (\ref{CA}), we have introduced, for arbitrary $x, y \in \mathbb{R}$ the $\mathcal{B}$-function
\begin{equation}
\mathcal{B}(x,y)=\frac{1}{2}\left(\frac{x+y}{|x+y|}-\frac{x-y}{|x-y|}\right) \label{d1}
\end{equation}
which returns $\text{sign}(y)$ if $-y < x < y$, zero if $|x|>y$ and $\text{sign}(y)/2$ if $|x|=y$. The $\mathcal{B}$-function is plotted in Fig. \ref{box} and has the form of a rectangular function whose thickness is controlled by the value of $y$. Clearly, if $x$ is an integer and $0< y \le 1/2$, the $\mathcal{B}$-function behaves as a Kronecker delta. 
We note, however, the following important fact that is a key to understand some aspects of the extension that we present below: \emph{Although the spatiotemporal dynamics of Eq. (\ref{CA}) fully takes places on the integers in $S \equiv \{0,\ldots, p-1\}$, the initial condition $x_{0}^{j}$ for $j$ arbitrary can be any real number such that $n-\sum_{k=-r}^{l}p^{k+r}x_{0}^{j+k} \ne \pm \frac{m}{2}$ where $m \in [1,p^{l+r+1}]$ is an integer}. (This is, of course, automatically satisfied if all $x_{0}^{j}$ are integers in $S$.) Therefore, \emph{there exists open continuous intervals of real numbers which are mapped to the same resulting spatiotemporal evolution on the integers}.

\begin{figure*}
\includegraphics[width=0.5 \textwidth, angle=270]{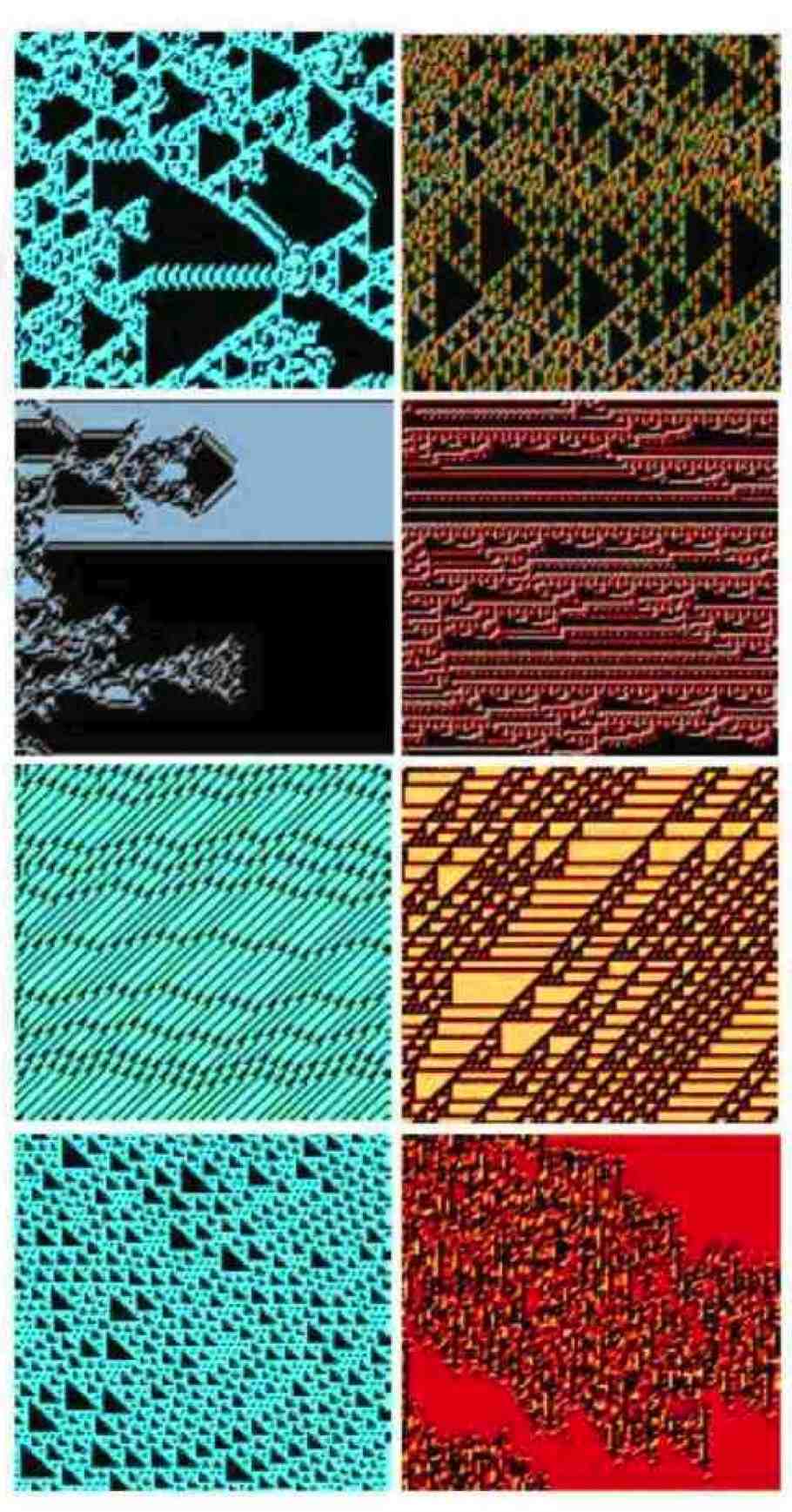}
\caption{\scriptsize{Spatiotemporal evolution of some 1D CA rules obtained from Eq. (\ref{CA}) for an arbitrary initial condition. From top to bottom and right to left the following rules are shown: $^1 110 _2 ^1$, $^2 29649 _2 ^1$, $^2 52T _2 ^2$, $^3 88T _2 ^3$, $^3 9334T _3 ^4$, $^2 51649 _2 ^1$, $^0 8322 _3 ^1$, $^1 93340T_5 ^1$. (In rules with totalistic code, Eq. (\ref{CAT}) is used instead). In each panel, time flows from top to bottom and space spans along the horizontal direction. Shown is a window $100\times 100$ in each case \cite{comphys}.}} \label{cas}  
\end{figure*}

Any CA described by Eq. (\ref{CA}) can be labelled with a code $^{l}R^{r}_{p}$ \cite{VGM1}, where $R$ is an integer $R \in [0,p^{p^{l+r+1}}-1]$ given by
\begin{equation}
R \equiv \sum_{n=0}^{p^{r+l+1}-1}a_{n}p^{n}. \label{RWolf}
\end{equation}
from which the coefficients of the CA rule $a_{n}$ can be obtained by means of the digit function $\mathbf{d}_{p}(n,R)$ (see \cite{CHAOSOLFRAC, PHYSAFRAC, QUANTUM, VGM4} for a detailed discussion of this function) as \cite{semipredo}
\begin{equation}
a_{n}=\mathbf{d}_{p}(n,R)=\left \lfloor \frac{R}{p^{n}} \right \rfloor-p\left \lfloor \frac{R}{p^{n+1}} \right \rfloor    \label{cucuAreal}
\end{equation}
where $\left \lfloor \ldots \right \rfloor$ denotes the lower-closest-integer (floor) function. 

Totalistic CA, whose output value depend on the sum over neighborhood values, are a subset of all ones described by Eq. (\ref{CA}) and the following simpler map \cite{VGM1} can alternatively be used
\begin{equation}
x_{t+1}^{i}=\sum_{s=0}^{(l+r+1)(p-1)}\sigma_{s}\mathcal{B}\left(s-\sum_{k=-r}^{l}x_{t}^{i+k}, \frac{1}{2}\right) \label{CAT}
\end{equation}
where $\sigma_{s} \in S$ specify the totalistic CA rule. The latter has code $^{l}RT_{p}^{r}$, with $R=\sum_{s=0}^{\rho(p-1)}a_{s}p^{s}$ (the $T$ is just a character that informs that the rule is totalistic).  
In Fig. \ref{cas} the spatiotemporal evolution of some CAs obtained from Eqs. (\ref{CA}) and (\ref{CAT}) is shown for an arbitrary initial condition. 

\begin{figure*}
\includegraphics[width=0.75 \textwidth]{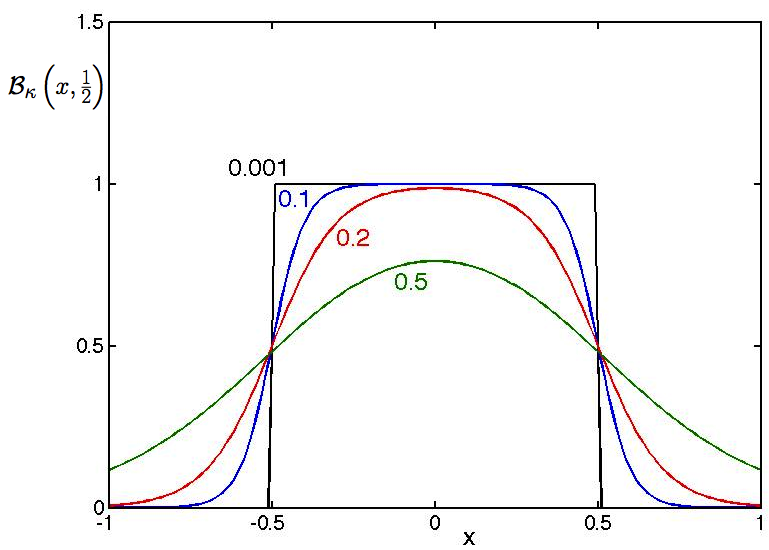}
\caption{\scriptsize{The function $\mathcal{B}_{\kappa}\left(x, \frac{1}{2} \right)$ for the values of $\kappa$ indicated in the figure.}} \label{generbo}  
\end{figure*}

We note that if $H(x)$ denotes the Heaviside step function we have
\begin{equation}
\mathcal{B}(x,y)=H(x+y)-H(x-y)
\end{equation}
and, therefore, if we define
\begin{equation}
\mathcal{B}_{\kappa}(x,y)\equiv \frac{1}{2}\left( 
\tanh\left(\frac{x+y}{\kappa} \right)-\tanh\left(\frac{x-y}{\kappa} \right)
\right) \label{bkappa}
\end{equation}
we see that, we have, as well
\begin{equation}
\lim_{\kappa \to 0} \mathcal{B}_{\kappa}(x,y)=H(x+y)-H(x-y)=\mathcal{B}(x,y)
\end{equation}
We thus see that $\mathcal{B}_{\kappa}(x,y)$ gradually smooths $\mathcal{B}(x,y)$ as $\kappa$ is increased. In Fig. \ref{generbo} this effect of changing $\kappa$ on the function $\mathcal{B}_{\kappa}(x,y)$ is shown.  

The following mathematical properties of $\mathcal{B}_{\kappa}(x,y)$, of fundamental importance to our approach, are easily established. The $\mathcal{B}_{\kappa}(x,y)$ function is an even function of its first argument and an odd function of its second argument
\begin{eqnarray}
\mathcal{B}_{\kappa}(-x,y)&=&\mathcal{B}_{\kappa}(x,y) \label{pari1} \\
\mathcal{B}_{\kappa}(x,-y)&=&-\mathcal{B}_{\kappa}(x,y) \label{pari2}
\end{eqnarray}
We also have the \emph{splitting property}. For $x, y, z$ any real numbers, we have
\begin{eqnarray}
\mathcal{B}_{\kappa}(x,y+z)&=&\mathcal{B}_{\kappa}(x+y,z)+\mathcal{B}_{\kappa}(x-z,y) \label{splito} 
\end{eqnarray}
This property can be easily proved by using the definition. We have
\begin{eqnarray}
&&\mathcal{B}_{\kappa}(x+y,z)+\mathcal{B}_{\kappa}(x-z,y)=\frac{1}{2}\left(\tanh\left(\frac{x+y+z}{\kappa}\right)-\tanh\left(\frac{x+y-z}{\kappa}\right)\right)+\nonumber \\
&&+\frac{1}{2}\left(\tanh\left(\frac{x-z+y}{\kappa}\right)-\tanh\left(\frac{x-z-y}{\kappa}\right)\right)=\frac{1}{2}\left(\tanh\left(\frac{x+y+z}{\kappa}\right)-\tanh\left(\frac{x-y-z}{\kappa}\right)\right) \nonumber \\
&&=\mathcal{B}_{\kappa}(x,y+z) \nonumber
\end{eqnarray}

We call the following property the \emph{block coalescence} property
\begin{eqnarray}
\sum_{k=0}^{n-1}\mathcal{B}_{\kappa}(x-2ky,y)&=&\mathcal{B}_{\kappa}(x-(n-1)y,ny) \label{induin} 
\end{eqnarray}
This result is easily proved by induction. For $n=1$, it is obviously valid. Let us assume it also valid for $n$ terms. Then, for $n+1$ terms we have
\begin{eqnarray}
\sum_{k=0}^{n}\mathcal{B}_{\kappa}(x-2ky,y)&=&\mathcal{B}_{\kappa}(x-2ny,y)+\sum_{k=0}^{n-1}\mathcal{B}_{\kappa}(x-2ky,y) \nonumber \\
&=&\mathcal{B}_{\kappa}(x-2ny,y)+\mathcal{B}_{\kappa}(x-(n-1)y,ny) \nonumber \\
&=&\mathcal{B}_{\kappa}(x-ny,(n+1)y)-\mathcal{B}_{\kappa}(x-(n-1)y,ny)+\mathcal{B}_{\kappa}(x-(n-1)y,ny) \nonumber \\
&=&\mathcal{B}_{\kappa}(x-ny,(n+1)y) \nonumber
\end{eqnarray} 
where the splitting property, Eq. (\ref{splito}) has been used. Thus, the induction principle establishes the validity of Eq. (\ref{induin}). 

The following asymptotic regimes are now noticed. For $\kappa \to 0$
\begin{equation}
\mathcal{B}_{\kappa}(x,y) \sim e^{-\frac{2(|x|-|y|)}{\kappa}}  \text{sign}(y) \label{asinto1}
\end{equation}
or, if the limit is strictly taken $\mathcal{B}_{\kappa}(x,y)  \to \mathcal{B}(x,y)$, as mentioned above. For $\kappa$ sufficiently large so that $\frac{|x|+|y|}{\kappa} < \frac{\pi}{2}$ for all relevant $x$ and $y$, the Taylor series of the hyperbolic tangents in Eq. (\ref{bkappa}) converge, and we have
\begin{eqnarray}
\mathcal{B}_{\kappa}(x,y)&=&\frac{1}{2}\sum_{m=1}^{\infty}\frac{2^{2m}(2^{2m}-1)B_{2m}}{(2m)!\kappa^{2m-1}}\left[\left(x+y \right)^{2m-1}-\left(x-y \right)^{2m-1}\right] \nonumber \\
&=&  \frac{1}{2}\sum_{m=1}^{\infty}\frac{2^{2m}(2^{2m}-1)B_{2m}}{(2m)!}
\sum_{h=0}^{2m-1}{2m-1 \choose h}\frac{x^{2m-1-h}y^{h}(1-(-1)^{h})}{\kappa^{2m-1}} \nonumber \\
&=&  \sum_{m=1}^{\infty}\frac{2^{2m}(2^{2m}-1)B_{2m}}{(2m)!}
\sum_{h=1}^{m}{2m-1 \choose 2h-1}\frac{x^{2(m-h)}y^{2h-1}}{\kappa^{2m-1}}  \label{Bernoul}
\end{eqnarray}
where the $B_{2m}$ are the even Bernoulli numbers: $B_{0}=1$, $B_{2}=\frac{1}{6}$, $B_{4}=-\frac{1}{30}$ $B_{6}=\frac{1}{42}$, etc. For $\kappa$ asymptotically large, this latter expression becomes, 
\begin{equation}
\mathcal{B}_{\kappa}(x,y) \sim  \frac{y}{\kappa}\left[1-\frac{3x^{2}+y^{2}}{3\kappa^{2}} \right]  \label{asinto2}
\end{equation}

By regarding all above properties, we can now embed the universal map, Eq. (\ref{CA}) (whose dynamics develops over the set $S$ of the integers $\in [0,p-1]$) in the reals within the same interval $[0,p-1]$ if we replace the $\mathcal{B}$-function in Eq. (\ref{CA}) by $\mathcal{B}_{\kappa}(x,y)$ given by Eq. (\ref{bkappa}). Then, we obtain the map
\begin{equation}
x_{t+1}^{j}=\sum_{n=0}^{p^{r+l+1}-1}a_{n}\mathcal{B}_{\kappa}\left(n-\sum_{k=-r}^{l}p^{k+r}x_{t}^{j+k},\frac{1}{2}\right) \label{CAfuz}
\end{equation} 
where now $x_{t}^{j} \in \mathbb{R}$. For any non-vanishing value $\kappa \in \mathbb{R}$ finite, we refer to Eq. (\ref{CAfuz}) as themap for RDCA. In the limit $\kappa \to 0$, Eq. (\ref{CAfuz}) reduces to Eq. (\ref{CA}) and we call this limit the \emph{CA limit} of the map. 

Since a CA rule is labeled by the code $^{l}R^{r}_{p}$, we can now label an RDCA rule described by Eq. (\ref{CAfuz}) as $^{l}R^{r}_{p, \kappa}$. We then have
\begin{equation}
\lim_{\kappa \to 0}\ ^{l}R^{r}_{p, \kappa}=\ ^{l}R^{r}_{p}
\end{equation}

\begin{figure*}
\includegraphics[width=0.7 \textwidth]{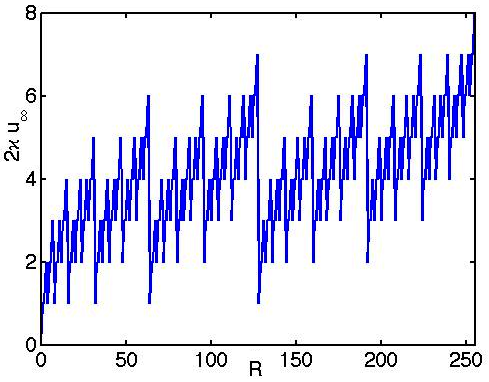}
\caption{\scriptsize{The value of the fixed point $2\kappa u_{\infty}$ as a function of $R$ calculated from Eq. (\ref{theglobalcrap}) in the asymptotic limit $\kappa \to \infty$ for $l=r=1$, $p=2$.}} \label{uinfi}  
\end{figure*}

Because of the asymptotic properties of the $\mathcal{B}_{\kappa}$-function above, we have, from Eq. (\ref{asinto1}) that for $\kappa \sim 0$, Eq. (\ref{CAfuz}) is asymptotically equivalent to
\begin{equation}
x_{t+1}^{j}=\sum_{n=0}^{p^{r+l+1}-1}a_{n}\exp{\left(\frac{1-2\left|n-\sum_{k=-r}^{l}p^{k+r}x_{t}^{j+k}  \right|}{\kappa}\right)} \qquad \qquad (\kappa \to 0) \label{Casin1}
\end{equation}
From Eq. (\ref{Bernoul}), if $\kappa > \frac{n_{max}+\frac{1}{2}}{\pi}$, where $n_{max} \in [0,p^{l+r+1}-1]$ is the maximum $n$ such that $a_{n}$ is nonzero, Eq. (\ref{CAfuz}) is \emph{equivalent} to
\begin{equation}
x_{t+1}^{j}=\sum_{n=0}^{p^{r+l+1}-1}\sum_{m=1}^{\infty}\sum_{h=1}^{m}a_{n}\frac{2^{2(m-h)+1}(2^{2m}-1)B_{2m}}{(2m)!\kappa^{2m-1}}
{2m-1 \choose 2h-1}\left(n-\sum_{k=-r}^{l}p^{k+r}x_{t}^{j+k}\right)^{2(m-h)} \label{CABern}
\end{equation} 
Finally, if $\kappa >> \frac{n_{max}+\frac{1}{2}}{\pi}$, we can keep only the dominant terms in Eq. (\ref{CABern}) and Eq. (\ref{CAfuz}) becomes asymptotically equal to
\begin{equation}
x_{t+1}^{j}=\sum_{n=0}^{p^{r+l+1}-1}\frac{a_{n}}{2\kappa}\left[1-\frac{1}{\kappa^{2}}\left(n-\sum_{k=-r}^{l}p^{k+r}x_{t}^{j+k}\right)^{2}-\frac{1}{12\kappa^{2}} \right] \label{Casin2}
\end{equation} 
In the limit $\kappa$ very large, the map becomes independent of the location $j$ and sets  any arbitrary initial condition to a global homogeneous fixed point 
\begin{equation}
u_{\infty}=\frac{1}{2\kappa}\sum_{n=0}^{p^{l+r+1}-1}a_{n} \label{theglobalcrap}
\end{equation}
In Fig. \ref{uinfi}, the quantity $2\kappa u_{\infty}=\sum_{n=0}^{p^{l+r+1}-1}a_{n}$, as obtained from Eq. (\ref{theglobalcrap}), is plotted versus $R=\sum_{n=0}^{p^{l+r+1}-1}a_{n}p^{n}$ for all RDCA whose CA limit correspond to Wolfram's elementary 256 CA rules with $p=2$, $l=r=1$.

All above results are also easily obtained for totalistic CA rules by simply replacing the $\mathcal{B}$-function by the $\mathcal{B}_{\kappa}$-function in Eq. (\ref{CAT}) 
\begin{equation}
x_{t+1}^{i}=\sum_{s=0}^{(l+r+1)(p-1)}\sigma_{s}\mathcal{B}_{\kappa}\left(s-\sum_{k=-r}^{l}x_{t}^{i+k}, \frac{1}{2}\right) \label{CATwoman}
\end{equation}
and we can label any totalistic RDCA rule described by Eq. (\ref{CATwoman}) as $^{l}RT^{r}_{p, \kappa}$. We then have
\begin{equation}
\lim_{\kappa \to 0}\ ^{l}RT^{r}_{p, \kappa}=\ ^{l}RT^{r}_{p}
\end{equation}
and all asymptotic results are easily derived again from the asymptotic properties of the $\mathcal{B}_{\kappa}$-function.

\section{Spatiotemporal evolution of the RDCA map} \label{numerical}

As a result of the above development we expect that, in varying the control parameter $\kappa$, bifurcations should appear marking different regimes in which Eq. (\ref{CAfuz}) behaves as a deterministic CA  at low $\kappa$ (and accurately described by Eq. (\ref{CA})), to a coupled map lattice at intermediate $\kappa$ and, finally, to the global quiescent state as $\kappa \to \infty$, as predicted by Eq. (\ref{theglobalcrap}). Then we expect some nontrivial behavior for intermediate $\kappa$, in the crossover between the above predicted asymptotic limits.

To get insight in Eq. (\ref{CAfuz}) it is better first to study the evolution of the global quiescent state. Thus we consider the special initial condition for which $x_{0}^{j}=0$ $\forall j$. All CA rules for which $a_{0}=0$ map the global quiescent state to itself, fixing it.  We thus expect that for a RDCA with $a_{0}=0$ and $\kappa$ nonvanishing but small, a global homogeneous state $x_{\infty}^{j}=u_{\infty} \approx 0$ should be a \emph{stable fixed point} to global perturbations. 

Let us study this global homogeneous state for such RDCA rules in the CA limit. We have $a_{0}=0$ and we take $a_{1} \ne 0$. From Eq. (\ref{CAfuz}), the dynamics of a global homogeneous state $x_{t}^{j}=u_{t}$ ($\forall j$) is governed by the map
\begin{eqnarray}
u_{t+1}&=&\sum_{n=0}^{p^{l+r+1}-1}a_{n}\mathcal{B}_{\kappa}\left(n-u_{t}\frac{p^{l+r+1}-1}{p-1},\frac{1}{2}\right) \label{CAfuzUNIF}
\end{eqnarray}
where $u_{0}=0=x_{0}^{j}$ $\forall j$. If $\kappa \to 0$ then $u_{\infty}=u_{0}=a_{0}=0$. If $\kappa$ is finite and small but nonvanishing, $u_{\infty}$ does not coincide with the quiescent state. To see this, note first that by taking $u_{0}=0$ in Eq. (\ref{CAfuzUNIF})
\begin{eqnarray}
u_{1}&=&\sum_{n=0}^{p^{l+r+1}-1}a_{n}\mathcal{B}_{\kappa}\left(n,\frac{1}{2}\right) \sim  
a_{1}e^{-\frac{1}{\kappa}} \ne 0 \label{CAfuzUNIF2}
\end{eqnarray}
where Eq. (\ref{asinto1}) has been used. Indeed, if $u_{t}$ is sufficiently small $\forall t$, Eq. (\ref{CAfuzUNIF}) is approximated in the CA limit by the transcendental map
\begin{equation}
u_{t+1}= a_{1}e^{-\frac{2}{\kappa}\left(\frac{1}{2}-u_{t}\frac{p^{l+r+1}-1}{p-1} \right)}=f(u_{t})
\end{equation}
from which the fixed point $u_{\infty}$ is obtained 
\begin{eqnarray}
u_{\infty}&=& a_{1}e^{-\frac{2}{\kappa}\left(\frac{1}{2}-u_{\infty}\frac{p^{l+r+1}-1}{p-1} \right)}=f(u_{\infty}) \ne 0 \label{CAfp}
\end{eqnarray} 
Such fixed point becomes unstable if
\begin{equation}
\left |\left.\frac{df(u_{t})}{du_{t}} \right|_{u_{\infty}} \right| > 1
\end{equation} 
i.e. if
\begin{equation}
\frac{\kappa}{u_{\infty}} < 2\frac{p^{l+r+1}-1}{p-1} 
\end{equation}
Thus, we have a bifurcation line at $\kappa_{L} \equiv 2\frac{p^{l+r+1}-1}{p-1}u_{\infty}$, which happens for the value $\kappa_{L}$ which is a solution of the transcendental equation
\begin{equation}
\kappa_{L}=2\frac{p^{l+r+1}-1}{p-1}a_{1}e^{-\frac{1-\kappa_{L}}{\kappa_{L}}} \label{bifuka}
\end{equation}
\emph{This expression is valid for all RDCA rules in rule space for which $a_{0}=0$ and $a_{1} \ne 0$.} For given $p$, $l$, $r$ values, there are a total of $p^{p^{l+r+1}}$ rules out of which $(p-1)p^{p^{l+r+1}-2}$ have the bifurcation governed by Eq. (\ref{bifuka}). For the RDCA rules constructed out of Wolfram's 256 elementary CA, i.e. those for which $p=2$ and $l=r=1$, there are 64 such rules. Those rules are the ones for which the Wolfram code $R$ is divisible by $2$ but not by $4$. Most interesting rules as rule 30 (a random number generator) and rule 110 (a universal Turing machine) belong to this set. We find, for all these rules, by setting  $p=2$ and $l=r=a_{1}=1$ in Eq. (\ref{bifuka}) the value $\kappa_{L}=0.18838296\ldots$.

\begin{figure*}
\includegraphics[width=0.7 \textwidth]{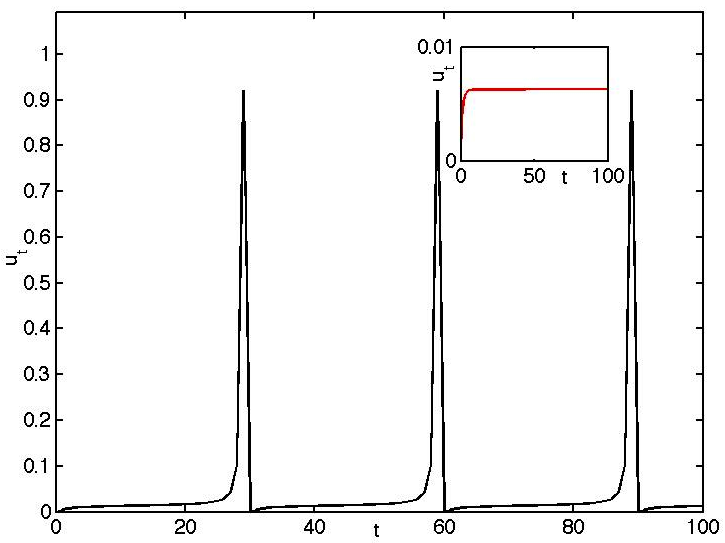}
\caption{\scriptsize{Evolution of $u_{t}$ obtained from Eq. (\ref{CAfuzUNIF}) for the rule $^{1}2^{1}_{2, \kappa}$ for $\kappa=0.19$. In the inset $\kappa=0.18$.}} \label{spike}  
\end{figure*}

We thus find that for $\kappa$ small finite and nonvanishing the global homogeneous fixed point loses its stability at $\kappa=\kappa_{L}$. Just beyond $\kappa_{L}$ oscillations in the form of spikes appear. When starting from the global quiescent state $u_{0}=0$, $u_{t}$ is small for a certain number of discrete time steps for which $u_{t} < u_{\infty}$, i.e. all those $t$ for which
\begin{equation}
a_{1}e^{-\frac{2}{\kappa}\left(\frac{1}{2}-u_{t}\frac{p^{l+r+1}-1}{p-1} \right)} < u_{\infty}
\end{equation}
However, since the fixed point is indeed repelling, $u_{t}$ grows exponentially as soon as 
$u_{t}\frac{p^{l+r+1}-1}{p-1}>\frac{1}{2}$. But $u_{t}$ cannot blow up to infinity because $\mathcal{B}_{\kappa}\left(n-u_{t}\frac{p^{l+r+1}-1}{p-1},\frac{1}{2}\right)=0$ for all finite $n$ in Eq. (\ref{CAfuzUNIF}) as soon as $u_{t}$ becomes sufficiently large. Thus, $u_{t}$ grows to a maximum value $u_{max} \approx a_{n_{max}}$ (where $n_{max}$ is the maximum $n$ such that $a_{n}$ is nonzero) after which it is suddenly reset to zero again. This behavior for the homogeneous mode reminds the bursting oscillations close to a homoclinic trajectory described in \cite{Afraimovich} for lattice systems. What is new here is that this behavior is approached from the CA limit in a way that a bifurcation of a universal character is discovered and exactly determined. Indeed, what this analysis reveals is that homoclinic chaos can be understood from a CA dynamical point of view (where the origin and the maximum distance attained within the loop of the homoclinic trajectory, as well as the shape of the oscillations, are related to the discrete states of a CA, which are connected through the homoclinic trajectory). We strongly believe that the underlying CA-like behavior of discrete systems is responsible for the ubiquity of excitable dynamics in nature \cite{Wolfram}.

In Fig. \ref{spike} the evolution of $u_{t}$, as obtained from Eq. (\ref{CAfuzUNIF}) for the rule $^{1}2^{1}_{2, \kappa}$ for $\kappa=0.19$ and initial condition $u_{0}=0$, is plotted. The bifurcation takes place at the value $\kappa_{L}=0.1884\ldots$ predicted by the above analysis. For $\kappa < \kappa_{L}$ a fixed point with numerical value given by Eq. (\ref{CAfp}) is attained, as shown in the inset of the figure for $\kappa=0.18$. However, as soon as $\kappa > \kappa_{L}$ the fixed point loses stability to relaxation oscillations (spikes).

\begin{figure*}
\includegraphics[width=1.0 \textwidth]{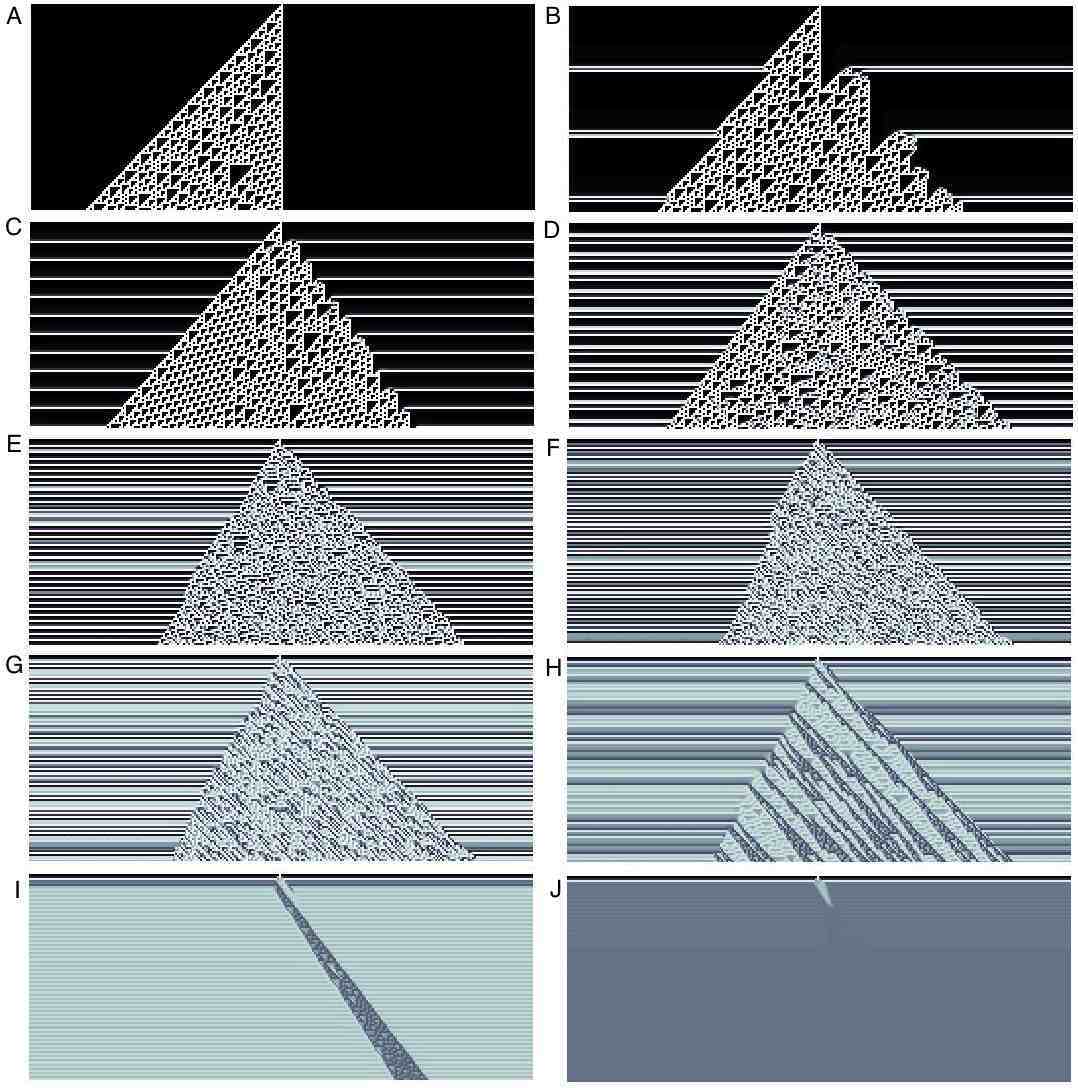}
\caption{\scriptsize{Spatiotemporal evolution of the map Eq. (\ref{CAfuz}) starting from a single site with value 1 surrounded by sites with zero value, for $p=2$, $l=r=1$, $(a_{0},a_{1},...a_{7})=(0,1,1,1,0,1,1,0)$, (i.e. $R=110$) and: $\kappa=0.18$ (A), $\kappa=0.19$ (B), $\kappa=0.20$ (C), $\kappa=0.22$ (D), $\kappa=0.30$ (E), $\kappa=0.40$ (F), $\kappa=0.50$ (G), $\kappa=0.70$ (H), $\kappa=0.80$ (I), $\kappa=1.00$ (J). In each panel, time flows from top to bottom. Shown is a window $256\times 100$ in each case.}} \label{kala}  
\end{figure*}

We now study the impact of this result when inhomogeneities are considered and, thus, we  investigate the full dynamics given by Eq. (\ref{CAfuz}). It proves useful to concentrate in the case in which the inhomogeneity is initially localized in a tiny region in space. Note that, in such case, for RDCA rules for which $a_{0}=0$ we can more neatly monitor the state of the homogeneous and the inhomogeneous phases and their eventual interaction, as a function of $\kappa$, because the inhomogeneity cannot propagate faster than $\max{(l,r)}$ sites per discrete time step.

We have performed numerical calculations of Eq. (\ref{CAfuz}) for the RDCA rule $^{1}110^{1}_{2, \kappa}$ for a simple initial condition in which a single seed with value `1' is surrounded by a phase in the quiescent state (with value `0'). We take periodic boundary conditions.
In the limit $\kappa \to 0$ this RDCA coincides with Wolfram's 110 CA rule. The results are shown in Fig. \ref{kala} for different increasing values of the parameter $\kappa$. We find that for $0 \le \kappa < 0.188$ (as in Fig. \ref{kala}A) the behavior of $^{1}110^{1}_{2, \kappa}$ is similar to Wolfram's CA rule 110. The only difference is that the `on' state is a real number, slightly lower than one and the `off' state is close to zero, but nonvanishing.  The evolution of these two states is the same as the asymptotic one obtained in the limit $\kappa \to 0$: The initial condition is a single seed and the inhomogeneity spreads asymmetrically forming a triangular structure which emits complex coherent structures to its interior. However, as $\kappa$ is increased, we find that at $\kappa \approx 0.1884$ the bifurcation described above takes place and we observe (Fig. \ref{kala}B) that the quiescent state outside the triangle of propagation of the inhomogeneity loses stability and periodically generates spikes that interact with the inhomogeneous phase introducing `defects'. We make the following observation of general validity before and just after the bifurcation: \emph{Within the inhomogeneity, the dynamics of the most significant digit of $x_{t}^{j}$ is governed by the CA to which the RDCA reduces in the CA limit while the less significant digits reflect the action of the smoothing due to a finite non-vanishing $\kappa$ value.} We note that the propagation for the most significant digit of $x_{t}^{j}$ is CA-like, exactly as in A, until the first spike happens. Then, new inhomogeneities are injected that interact with the previous one. As $\kappa$ is increased beyond the bifurcation, Fig \ref{kala}C, we observe that the frequency of the spikes also increases and that defects are being injected at a higher pace on the triangular structure. We note that there are asymmetric edge effects introduced by the border of the triangular structure: The static border is felt at a finite distance deep within the homogeneous phase and does not directly intersect the spike, while the border propagating a constant velocity directly intersects the spike being produced in the homogeneous phase. As a consequence of this asymmetry, the triangular inhomogeneous structure tends to rotate counterclockwise as $\kappa$ is increased and more and more defects are being injected. The propagation within the triangular structure is still CA-like, although the triangle of propagation of the inhomogeneities grows asymmetrically through this injection of defects. For larger $\kappa$, the injection of defects happens at such a larger pace that it becomes chaotic and the smaller homogeneous regions contained in the small triangles propagating within the inhomogeneous triangle are also affected and a disruption of the CA-like behavior takes place within the inhomogeneity. In panels E to H of Fig. \ref{kala} this trend continues, but now the dynamics of relaxation to a global homogeneous fixed point is being approached chaotically, which implies a stronger mixing of the inhomogeneities. As a consequence of the interaction between the triangular structure and the chaotic homogeneous phase, the border of the triangular structure becomes turbulent. As $\kappa \to \infty$ (not shown) a global quiescent state is approached, as predicted by Eq. (\ref{theglobalcrap}). Indeed, for $\kappa= 1$ (panel J) a global homogeneous fixed point is already found.

\begin{figure*}
\includegraphics[width=1.0 \textwidth]{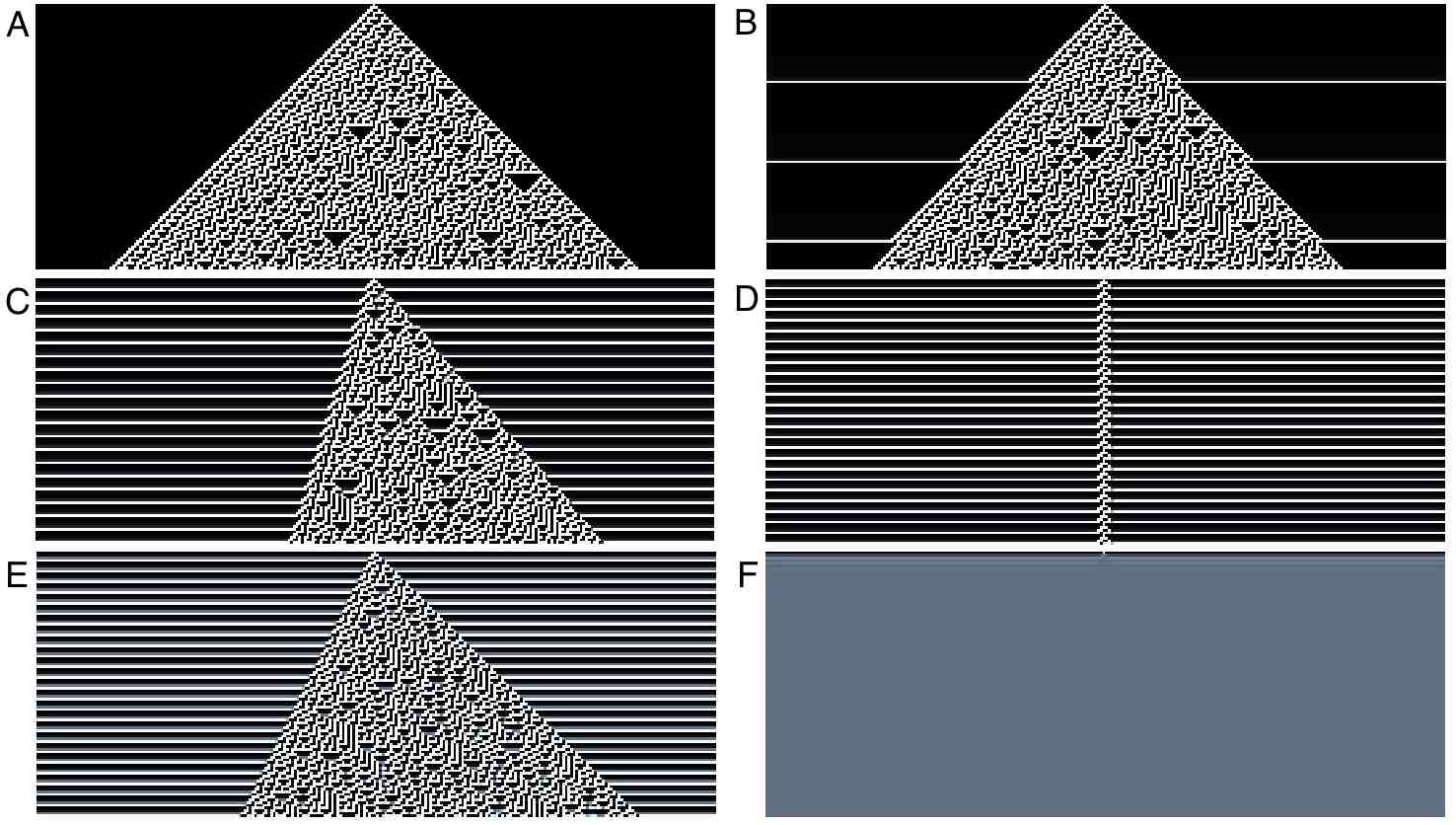}
\caption{\scriptsize{Spatiotemporal evolution of the map Eq. (\ref{CAfuz}) starting from a single site with value 1 surrounded by sites with zero value, for $p=2$, $l=r=1$, $(a_{0},a_{1},...a_{7})=(0,1,1,1,1,0,0,0)$, (i.e. $R=30$) and: $\kappa=0.18$ (A), $\kappa=0.19$ (B), $\kappa=0.25$ (C), $\kappa=0.3$ (D), $\kappa=0.4$ (E), $\kappa=2.0$ (F). In each panel, time flows from top to bottom. Shown is a window $256\times 100$ in each case.}} \label{kalb}  
\end{figure*}

The behavior for intermediate values of $\kappa$ is highly nontrivial and one cannot simply generally conclude that the dynamics becomes `more turbulent' as $\kappa$ is increased. An example of this strikingly counterintuitive behavior is observed in Fig. \ref{kalb} where the spatiotemporal evolution of RDCA rule $^{1}30^{1}_{2,\kappa}$ is shown. In the CA limit, this RDCA reproduces the famous Wolfram rule 30, which is known to be a random number generator and a hallmark of an autoplectic dynamical system \cite{Wolfram7}. Autoplectic dynamics as defined by Wolfram correspond to \emph{spatially extended systems that starting from a simple (non-random) initial condition are able to generate intrinsic randomness by themselves so that no regular pattern is discernible in their spatiotemporal evolution} \cite{semipredo}. In panel A in Fig. \ref{kalb} the evolution of the RDCA $^{1}30^{1}_{2,\kappa}$ for $\kappa=0.18$ is similar to Wolfram rule 30 (i.e. CA rule $^{1}30^{1}_{2}$ in our notation). As the bifurcation of the homogeneous mode described above is passed, spikes arise in the homogeneous phase which again inject defects in the triangular structure (panels B-E of Fig. \ref{kalb}). Strikingly, however, the chaotic inhomogeneous phase is destroyed by continuously increasing $\kappa$ in the range $0.25 < \kappa <0.3$. As a result, in panel $D$ we obtain a static coherent structure in place of the chaotic triangular structure. As $\kappa$ is further increased (panel $E$), the chaotic phase reappears.

\begin{figure*}
\includegraphics[width=0.8 \textwidth]{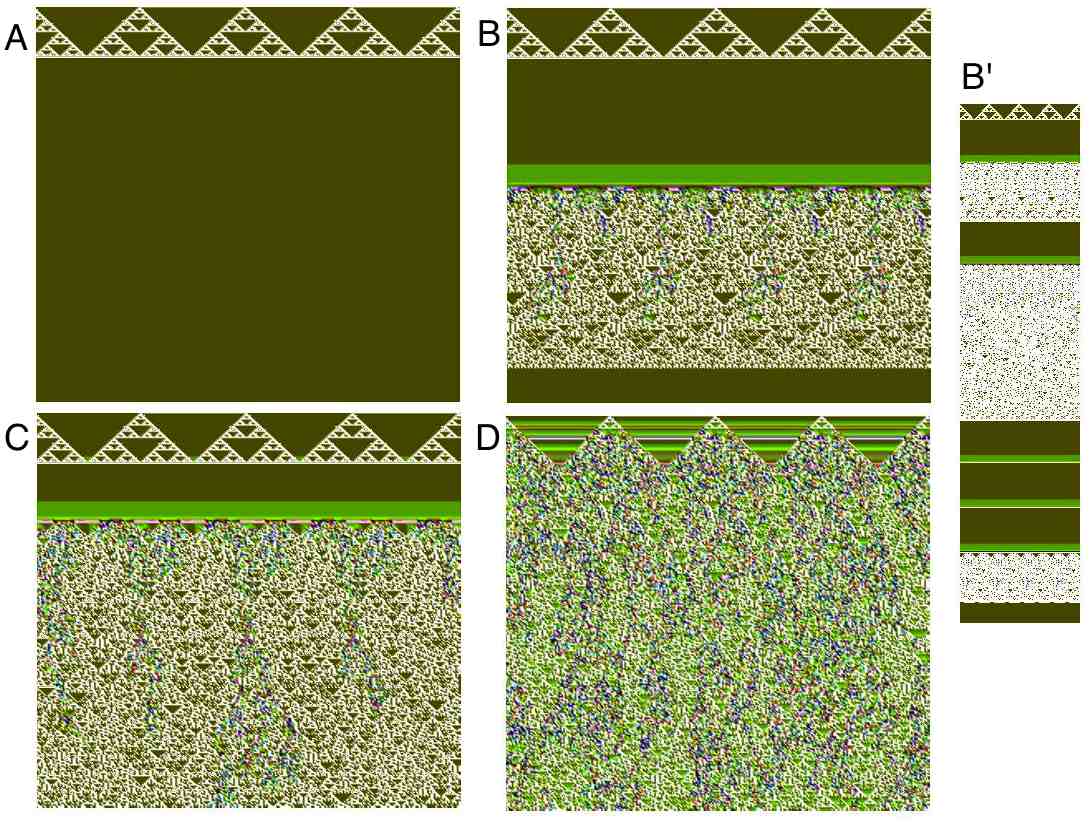}
\caption{\scriptsize{Spatiotemporal evolution of the map Eq. (\ref{CAfuz}) starting from an initial condition $x_{0}^{j}=\mathbf{d}_{2}(6,j)$ ($j\in [1,256]$), for $p=2$, $l=r=1$, $(a_{0},a_{1},...a_{7})=(0,1,0,1,1,0,1,0)$, (i.e. $R=90$) and: $\kappa=0.188$ (A), $\kappa=0.1891$ (B), $\kappa=0.1895$ (C) and $\kappa=0.2$ (D). Shown is a window $256\times 256$ in each case. In panel B' parameters are as in B, but a window $256 \times 1024$ is shown instead. Time flows from top to bottom in every panel.}} \label{kali}  
\end{figure*}

\begin{figure*}
\includegraphics[width=1.0 \textwidth]{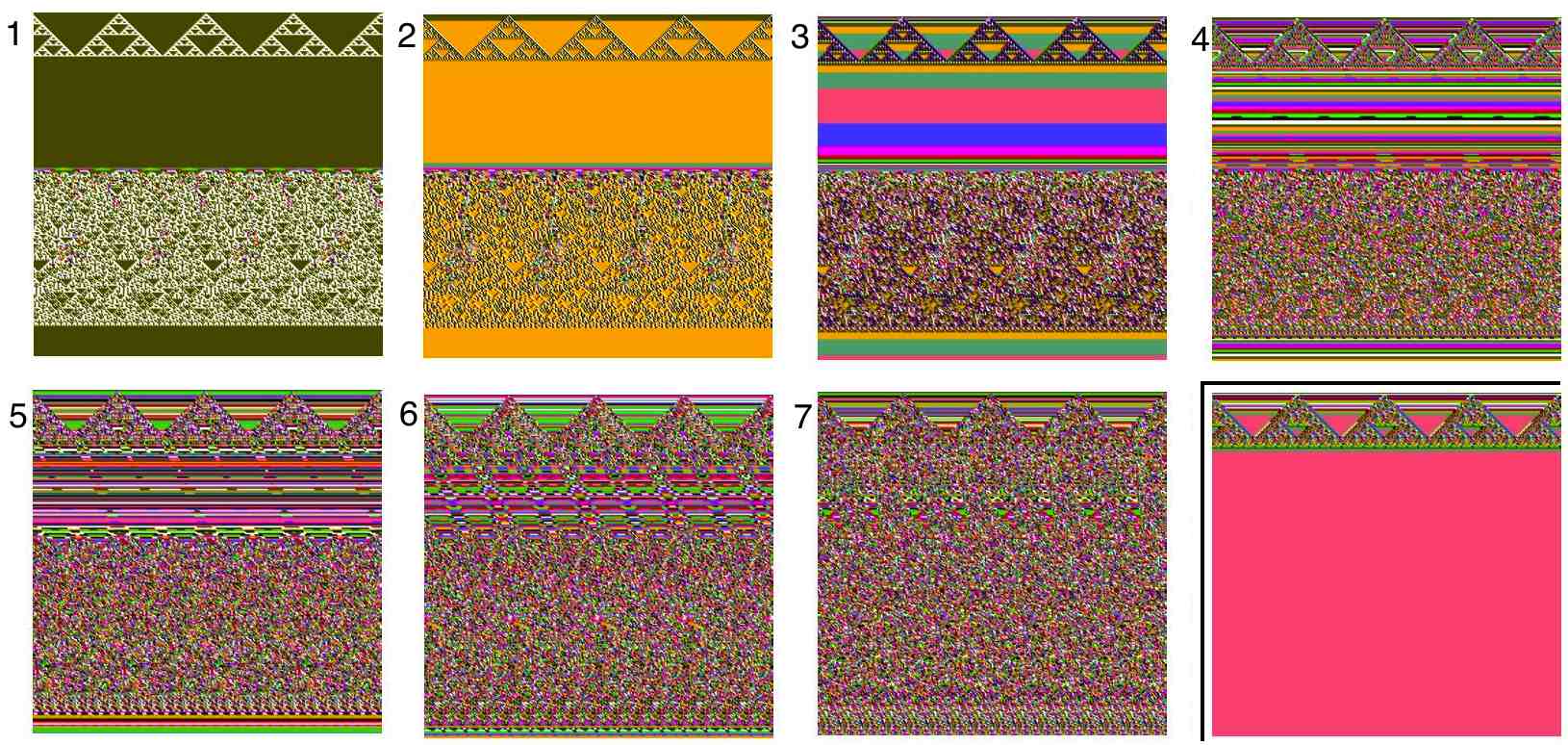}
\caption{\scriptsize{Spatiotemporal evolution of $\mathbf{d}_{10}(-k,x_{t}^{j})$ for $x_{t}^{j}$ provided by panel B in Fig. \ref{kali} and the values of $k$ indicated in the panels. In the bottom rightmost panel $\mathbf{d}_{10}(-7,x_{t}^{j})$ for $x_{t}^{j}$ taken from panel A in Fig. \ref{kali} is shown for comparison.}} \label{kalu}  
\end{figure*}

We can explain this unusual behavior with help of the mathematical treatment presented in this paper. For RDCA rule $^{1}30^{1}_{2,\kappa}$, we have $(a_{0},a_{1},...a_{7})=(0,1,1,1,1,0,0,0)$ and Eq. (\ref{CAfuz}) reduces to
\begin{eqnarray}
x_{t+1}^{j}&=&\sum_{n=0}^{p^{r+l+1}-1}a_{n}\mathcal{B}_{\kappa}\left(n-\sum_{k=-r}^{l}p^{k+r}x_{t}^{j+k},\frac{1}{2}\right) = \sum_{n=1}^{4}\mathcal{B}_{\kappa}\left(n-\sum_{k=-1}^{1}2^{k+1}x_{t}^{j+k},\frac{1}{2}\right) \nonumber \\
&=&\sum_{n'=0}^{3}\mathcal{B}_{\kappa}\left(n'+1-\sum_{k=-1}^{1}2^{k+1}x_{t}^{j+k},\frac{1}{2}\right) 
\end{eqnarray} 
which, by using Eq. (\ref{pari1}) and the block coalescence property, Eq. (\ref{induin}), with $x=-1+\sum_{k=-1}^{1}2^{k+1}x_{t}^{j+k}$ and $y=\frac{1}{2}$ reduces to
\begin{eqnarray}
x_{t+1}^{j}&=&\mathcal{B}_{\kappa}\left(\frac{5}{2}-\sum_{k=-1}^{1}2^{k+1}x_{t}^{j+k},2\right)
\end{eqnarray}
This means that $^{1}30^{1}_{2,\kappa}$ can be understood as a single block in terms of the $\mathcal{B}_{\kappa}$ function, which is centered at $5/2$ and has with 'thickness' 4. In the CA limit, such $\mathcal{B}_{\kappa}$ function outputs value 1 or almost 1 for $1\le \sum_{k=-1}^{1}2^{k+1}x_{t}^{j+k} \le 4$ and zero (or almost zero) for other values of $\sum_{k=-1}^{1}2^{k+1}x_{t}^{j+k}$. As $\kappa$ increases beyond the bifurcation, the block is smoothed and for $\kappa$ sufficiently large there are nonzero contributions of all values  $0\le \sum_{k=-1}^{1}2^{k+1}x_{t}^{j+k} \le 5$, i.e. there is a significant nonzero contribution for the neighborhood configurations '000' and '101' as well. This means that the behavior of CA rule $^{1}63^{1}_{2}$ is approximated. This latter rule is \emph{not} chaotic, but a Class 2 rule in Wolfram's CA classification. This leads to the seemingly destruction of the autoplectic character as observed in panel D of Fig. \ref{kalb} for a simple initial condition, and in the passage of $^{1}30^{1}_{2,\kappa}$ from a Class 3 rule to a Class 4 CA-like RDCA rule for a generic initial condition. The resulting RDCA rule at intermediate $\kappa$ is a mixture of the behaviors of Wolfram rules 30 and 63 in such a way that spatially extended ordered patches of rule 63 coexist with chaotic bursts caused by rule 30. This leads to spatiotemporal intermittency (results not shown). 

Another striking phenomenon found at intermediate $\kappa$ values past the bifurcation is a spatially homogeneous-spatially turbulent alternation found for rule $^{1}90^{1}_{2,\kappa}$. This phenomenon is also present in all RDCAs that, in the CA limit, perform the addition modulo $2$ of the neighborhood values \cite{VGM3} (Wolfram CA rules $^{1}60^{1}_{2}$, $^{1}102^{1}_{2}$ and $^{1}150^{1}_{2}$ are examples of this). For system sizes that are powers of the alphabet size ($p=2$ in this case) a global cancellation occurs at some specific time for certain initial conditions in the CA limit of the corresponding RDCAs. This is illustrated in Fig. \ref{kali} A, where the spatiotemporal evolution of $^{1}90^{1}_{2,\kappa}$ is shown in the CA limit ($\kappa=0.188$) for an initial condition $x_{0}^{j}=\mathbf{d}_{2}(6,j)$ ($j\in [1,256]$), i.e. for a ring of $256=2^{8}$ sites. After 32 time steps a global cancellation occurs and the system reaches a global homogeneous fixed point close to the quiescent state. As soon as the bifurcation predicted by our theory is crossed (see panel B, $\kappa=0.1891$ for the same initial condition), the global homogeneous fixed point is unstable and spiking oscillations set on. However, such spikes do not simply relax back to the global homogeneous state but to a turbulent state that lasts a finite time interval before the the global homogeneous state sets again. As shown in panel B' for a time span of 1024 time steps, the alternation between turbulence and global homogeneity is aperiodic, as are the time intervals occupied by the turbulent regime. As $\kappa$ is further increased, the time intervals between bursting events (panel C, $\kappa=0.1895$) are decreased and, for $\kappa=0.2$ the behavior is always turbulent.  

This intermittent behavior can be understood from the fact that addition modulo 2 leading to the global cancellation after 32 time steps is only exact in the CA limit $\kappa \to 0$ of $^{1}90^{1}_{2,\kappa}$. At nonvanishing $\kappa$ there are slight differences from the exact modulo 2 operation in the previous configurations that do not lead to an exact cancellation at $t=32$ (note that the hyperbolic tangents in the definition of $\mathcal{B}_{\kappa}$ are transcendental functions and, therefore, they can only \emph{approximate} the addition modulo 2 operation). These tiny differences, which are of order $10^{-5}$ to $10^{-7}$, constitute inhomogeneities that are masked by the homogeneous dynamics of the most significant digits of $x_{t}^{j}$. Thus, when the homogeneous mode is unstable, the inhomogeneities propagate from less to more significant digits and, at the spike, they come to occupy the most significant digits of the dynamics, which becomes again a CA dynamics till the next cancellation occurs. All these facts are revealed from an analysis of the spatiotemporal distribution of the different decimal digits of $x_{t}^{j}$
\begin{equation}
\mathbf{d}_{10}(-k,x_{t}^{j})=\left \lfloor 10^{k}x_{t}^{j} \right \rfloor-10\left \lfloor 10^{k-1}x_{t}^{j} \right \rfloor    \label{cucuAreal}
\end{equation}
for different values of $k$. The spatiotemporal evolution of $\mathbf{d}_{10}(-k,x_{t}^{j})$ for $k=1, 2,\ldots 7$ is shown as indicated in the panels for $\kappa=0.1891$ as in Fig. \ref{kali} B.
We see that the most significant digits $\mathbf{d}_{10}(-1,x_{t}^{j})$, $\mathbf{d}_{10}(-2,x_{t}^{j})$ and $\mathbf{d}_{10}(-3,x_{t}^{j})$ are homogeneous after the cancellation at $t=32$. However, the less significant digits are inhomogeneous and these tiny contributions become amplified at the spike that is produced as a consequence of the instability of the homogeneous mode. For $\kappa=0.18$ (before the bifurcation) even when $\kappa$ is finite and nonvanishing and the modulo 2 operation of the CA limit is here not exact, since the homogeneous fixed point is stable, the cancellation produced at $t=32$ on the most significant digits of $x_{t}^{j}$ drives the system to that homogeneous state. In the rightmost bottom panel of Fig. \ref{kalu} the digit of the seventh decimal place (after the decimal point) $\mathbf{d}_{10}(-7,x_{t}^{j})$ is shown and its perfect homogeneity shows that all considerations/observations above are based on exact calculations. The general conclusion of this analysis is: For $\kappa$ low, the most significant digits of $x_{t}^{j}$ evolve according to the CA dynamics obtained from the CA limit of the RDCA. When $\kappa$ is slightly larger than $\kappa_{L}$ given by Eq. (\ref{bifuka}) the homogeneous mode is unstable and any inhomogeneity, however slight, propagates within the less significant digits of the signal until a spike is produced in which case gets amplified, being `pumped' to the most significant digits of the $x_{t}^{j}$ and thus initiating `ex nihilo' another CA-like computation which lasts a finite time till the next cancellation. 

\section{Conclusions}

In this article we have presented an exact construction of a wide class of coupled map lattices from cellular automata that we have termed `real-valued deterministic cellular automata' and we have shown how it encompasses both CAs and infinite families of CMLs. Although the generalization presented in this article is certainly not unique and infinitely many other possibilities exist to smoothen or mollify the universal map for CA, we conjecture that any arbitrary CML is a RDCA or can be reasonably approximated by an RDCA of the type introduced here. The main feature of this class of dynamical systems, is the dependence on a continuous parameter $\kappa$ such that, when $\kappa \to 0$ (the CA limit), the dynamics of deterministic CA is exactly reproduced. In the opposite limit $\kappa \to \infty$ all RDCA rules derived here tend to a global homogeneous fixed point whose value has been exactly determined. 

We have uncovered a bifurcation, as $\kappa$ is increased, in which the homogeneous mode of CA dynamics loses stability to relaxation oscillations and we have exactly determined the location of the bifurcation in parameter space by means of a linear stability analysis. Then, we have also uncovered some nontrivial dynamical phenomena found in the interaction between the unstable homogeneous mode and the inhomogeneities. Most remarkably, a special kind of spatiotemporal intermittence has been found in which spatial homogeneity and turbulence alternate in time.  

The connection of the approach presented here to the ultradiscretization method \cite{Tokihiro1, Tokihiro2, Tokihiro3, Tokihiro4, Tokihiro5, Tokihiro6} and the algebro-geometric construction of fully discrete integrable systems \cite{Bialecki1, Bialecki2} would be interesting to gain additional insight in the complex interrelationships of CAs, CMLs and partial differential equations, and the subtleties involved in the discretization. Besides, other generalizations introducing mollifiers of $\mathcal{B}$-functions instead of $\mathcal{B}_{\kappa}$-functions might be interesting to explore the direct relationship of CAs to PDEs \cite{Omohundro}.

\bibliography{biblos}{}
\bibliographystyle{h-physrev3.bst}

\end{document}